\documentclass[a4paper,fleqn,usenatbib]{mnras}

\usepackage{newtxtext,newtxmath}

\usepackage[T1]{fontenc}
\usepackage{ae,aecompl}

\usepackage{graphicx}	
\usepackage{amsmath}	
\usepackage{amssymb}	

\title[KIC 8462852 Light Curve 2015.75 to 2018.18]{The KIC 8462852 Light Curve From 2015.75 to 2018.18 Shows a Variable Secular Decline}

\author[B. E. Schaefer et al.]{
Bradley E. Schaefer$^{1}$\thanks{E-mail: schaefer@lsu.edu},
Rory O. Bentley$^{1}$,
Tabetha S. Boyajian$^{1}$,
Phillip H. Coker$^{2}$,
\newauthor
Shawn Dvorak$^{2,3}$
Franky Dubois$^{2,4,5}$,
Emery Erdelyi$^{2,6}$,
Tyler Ellis$^{1}$,
\newauthor
Keith Graham$^{2}$,
Barbara G. Harris$^{2,7}$,
John E. Hall$^{2,8}$,
Robert James$^{2}$,
\newauthor
Steve J. Johnston$^{2,9}$,
Grant Kennedy$^{10}$,
Ludwig Logie$^{4,5}$,
Katherine M. Nugent$^{1}$,
\newauthor
Arto Oksanen$^{2,11}$,
John J. Ott$^{2,12}$,
Steve Rau$^{4,5}$,
Siegfried Vanaverbeke$^{4,5}$,
\newauthor
Rik van Lieshout$^{13}$,
and Mark Wyatt$^{13}$
\\
$^{1}$Department of Physics and Astronomy, Louisiana State University, Baton Rouge, Louisiana, 70820, USA\\
$^{2}$American Association of Variable Star Observers, 49 Bay State Road, Cambridge, Massachusetts, 02138, USA\\
$^{3}$Rolling Hills Observatory, Lake County, Florida, USA\\
$^{4}$AstroLAB IRIS Observatory, Verbrandemolenstraat, Ypres, Belgium\\
$^{5}$Vereniging voor Sterrenkunde, Werkgroep Veranderlijke Sterren, Belgium\\
$^{6}$KSE Observatory, Carlsbad, California, USA\\
$^{7}$Bar J Observatory, New Smyrna Beach, Florida, 32168, USA\\
$^{8}$Angel Peaks Observatory, Cotopaxi, Colorado, 81223, USA\\
$^{9}$Karen Observatory, 12 Handley Drive, Padgate, Warrington, Cheshire, WA2 0GN, UK\\
$^{10}$Department of Physics, University of Warwick, Coventry CV4 7AL, UK\\
$^{11}$Hankasalmi Observatory, Hankasalmi, Finland\\
$^{12}$Ott Observatory, Nederland Colorado, USA\\
$^{13}$Institute of Astronomy, University of Cambridge, Madingley Road, Cambridge CB3 0HA, UK\\
}

\date{Accepted XXX. Received YYY; in original form ZZZ}

\pubyear{2018}

\begin{document}
\label{firstpage}
\pagerange{\pageref{firstpage}--\pageref{lastpage}}
\maketitle

\begin{abstract}
The star KIC 8462852 (Boyajian's Star) displays both fast dips of up to 20\% on time scales of days, plus long-term secular fading by up to 19\% on time scales from a year to a century.  We report on CCD photometry of KIC 8462852 from 2015.75 to 2018.18, with 19,176 images making for 1,866 nightly magnitudes in BVRI.  Our light curves show a continuing secular decline (by 0.023$\pm$0.003 mags in the B-band) with three superposed dips with duration 120-180 days.  This demonstrates that there is a continuum of dip durations from a day to a century, so that the secular fading is seen to be by the same physical mechanism as the short-duration Kepler dips.  The BVRI light curves all have the same shape, with the slopes and amplitudes for VRI being systematically smaller than in the B-band by factors of 0.77$\pm$0.05, 0.50$\pm$0.05, and 0.31$\pm$0.05.  We rule out any hypothesis involving occultation of the primary star by any star, planet, solid body, or optically thick cloud.  But these ratios are the same as that expected for ordinary extinction by dust clouds.  This chromatic extinction implies dust particle sizes going down to $\sim$0.1 micron, suggesting that this dust will be rapidly blown away by stellar radiation pressure, so the dust clouds must have formed within months.  The modern infrared observations were taken at a time when there was at least 12.4\%$\pm$1.3\% dust coverage (as part of the secular dimming), and this is consistent with dimming originating in circumstellar dust.
\end{abstract}




\section{Introduction}

The star KIC 8462852 (Boyajian's Star) is a perfectly ordinary, isolated, and middle-aged F3 main sequence star in Cygnus.  By all precedence and theory, the star should be stable in brightness (to the millimagnitude level) on all time scales faster than many millions of years.  So it was a startling surprise when the star was stared at by the {\it Kepler} spacecraft and it was seen to undergo a chaotic series of dips in brightness, with time scales from one day to a month and more, some reaching in amplitude to more than 0.2 mag (Boyajian et al. 2016).  Early efforts to explain this phenomenon by relatively ordinary means all failed for various reasons; with the {\it Kepler} data being unimpeachable, with the star really being isolated and middle-aged, and with the complete lack of any infrared excess pointing to accretion disks or surrounding clouds.  Currently, an exceptionally wide variety of models have been proposed, but none of them have any positive evidence or any real plausibility (Wright \& Sigurdsson 2017).

Boyajian et al. (2018) report on the excellent photometry in B-, r'-, and i'-bands with hourly time resolution from 2017.33 to 2017.81, as taken with the fully-automated system of many telescopes distributed worldwide as part of the Las Cumbres Observatory.  This long series of observations was funded by a KickStarter program, as the only practical means of getting a very high cadence and a very good consistency by beating down the usual small systematic errors in photometry.  The stated goal was to find further {\it Kepler}-like dips in real time so as to trigger an intensive observing campaign with many telescopes on the ground and in space.  In this, the KickStarter program succeeded wonderfully, catching a small decline, triggering many target-of-opportunity programs, and watching in real time as a week-long dip of 1.7\% unfolded.  This dip centered at 2017.38 was named ``{\it Elsie}", rather than some cumbersome and forgettable long string of numbers.  After {\it Elsie}, the Las Cumbres photometry continued, catching three more dips centered on 2017.46 ({\it Celeste} at 1.3\%), 2017.60 ({\it Skara Brae} at 1.2\%), and 2017.69 ({\it Angkor} at 2.4\%).  A primary result is that the dips are much deeper in the blue than in the red, in exactly the manner as predicted for the occulter being composed of ordinary celestial dust (Boyajian et al. 2018; Deeg et al. 2018).  Further, no absorption lines were seen to grow during the dips, so any gaseous component of the dust clouds must be minimal or somehow hidden.  So apparently the dips are caused by dust clouds passing in front of the parent star.

Further, it was discovered with the Harvard plates from 1890 to 1989 that the Boyajian's Star faded by 0.193$\pm$0.030 mag from 1890 to 1989, with an apparent dip of 0.10 mag from 1900-1909 (Schaefer 2016)\footnote{Claims by Hippke et al. (2016) are refuted as being due to multiple technical errors, see for example https://www.centauri-dreams.org/?p=35666.}.  The existence of secular fading was soon confirmed with the quarterly {\it Kepler} full-frame images from 2009.3 to 2013.3, showing a decline at a rate of near 0.341$\pm$0.041\% per year for the first thousand days (for a drop of 0.9\%), then a fast drop by 2\% over $\sim$200 days, followed by a slow decline for the last 200 days (Montet \& Simon 2016).  The existence of variable secular declines has been further confirmed over 15 months with {\it Swift} data, {\it Spitzer} data, and {\it AstroLAB IRIS} data (Meng et al. 2018), plus confirmations from ground-based data over 27 months with {\it ASAS-SN} data, and from 2006 to 2017 with {\it ASAS} data (Simon et al. 2018).  The near-ultraviolet (centered at 2317\AA) flux faded by 3.5$\pm$1.0\% from 2011 to 2012 as seen with the {\it GALEX} satellite (Davenport et al. 2018).  With 835 Maria Mitchell plates, the light curve from 1922 to 1991 shows an average fading of 0.12$\pm$0.02 mag per century, confirming the measured decline from the Harvard plates (Castelaz \& Barker 2018).  So the existence of secular fading (with rates from 0.11 to 3.5 percent per year) has been confirmed by many groups, both spaceborne and ground-based, on time scales from a year to a century.

Starting with the first announcement of the peculiarities of KIC 8462852 in the middle of September 2015, we have started and continued making CCD photometric observations in many of the standard optical bands.  This set of magnitudes can be used to measure the long-term secular variations.

\section{Observations}

We have used twelve telescopes scattered around the world to get our CCD photometry of KIC 8462852.  A full journal of observations appears in Table 1.  The columns list the observers, the observatory, the telescope, the number of individual magnitudes, the number of nightly averages, the filters used, plus the four offsets in magnitudes to standardize between observers in the B-, V-, R-, and I-bands.

Our observations were taken with CCDs with a wide array of telescopes and cameras.  Each system has a different size of the field-of-view, ranging from 11 to 30 arc-minutes on a side.  With this, the choice of comparison stars must vary from observer to observer.  The number of comparison stars varied from just one, up to ten nearby stars used as an ensemble standard.  The magnitudes for all these comparison stars was taken from the charts of the {\it American Association of Variable Star Observers} ({\it AAVSO}) as part of their {\it AAVSO Photometric All-Sky Survey} ({\it APASS}), with the photometric system for B-band and V-band going back to the Landolt standards (Landolt 1992).  The full details of the specific comparison stars used and the adopted magnitudes are given for each CCD image in the AAVSO data base, which is publicly available on-line\footnote{https://www.aavso.org/data-download}.  We have made the usual checks for these comparison stars for variability, versus other nearby check stars of comparable magnitudes, from our own data, while further checks for variability of most of our comparison stars were made with the {\it Kepler} full-frame images (see Figures 1 and 5 of Montet \& Simon 2016), all showing that our chosen comparison stars have no significant variability.  The magnitudes for our particular {\it APASS} comparison stars have typical photometric uncertainties of $\approx$0.04 mag.  All our photometry is differential with respect to the comparison stars, so the usual small errors in the standards will result in a constant offset for each observer's light curve with respect to a light curve in some standard system.  With our photometry being made with each observer using a different set of comparison stars, it means that each reported light curve should be adjusted with a constant offset by several hundredths of a magnitude.  We cannot now know the exact true magnitudes for the comparison stars used, so this means that some unknown and small offset is required for each light curve so as to put it into a standard system.  In practice, when combining or comparing light curves, this means that we are allowed and required to add some small arbitrary constant offset to each light curve.  In practice, this is done by setting the offsets for each observer so as to minimize the overall scatter in the combined light curves.

Our observations were taken through standard filters.  But not all filters are identical.  More generally, everyone's filter+CCD spectral sensitivity is slightly different.  In principle, the differential photometry can be placed onto a standard system by the use of color-terms, of the form $C\times(B-V)$, being added to the instrumental magnitude.  The color term varies with the color of the star, here quantified as the color index $B-V$.  The size of the color term is quantified by the coefficient $C$, which is usually small.  For small color terms, the linear nature of the term is an excellent approximation of the exact correction from the full integral over all wavelengths.  The coefficient $C$ is never exactly zero and is not easy to measure with high accuracy.  Therefore, each individual observer will have a small constant offset for the magnitude of the target star (as compared to some standard photometric system) due to the particular relative colors of the target and comparison stars.  This offset is difficult to know with precision, and is only of importance when comparing light curves from different sources.  In practice, this is done by adding a small arbitrary constant to each observer's light curve, so as to minimize the scatter between the light curves.

\section{Photometry}

All our photometry is differential with respect to the nearby on-chip comparison stars.  The instrumental magnitudes for each star are measured with the usual aperture photometry.  For a case with instrumental magnitudes $m$ and standard magnitudes, say $V$, for the target and comparison stars, the standard magnitude of the target will be $V=V_{comp}+m-m_{comp}$.  There will be additional small terms added in for differential airmasses and colors of the two stars.  For purposes of constructing a long term light curve over many nights, in principle, we simply keep using the same comparison stars and magnitudes, and we should get the light curve with no offsets from night-to-night.

The statistical errors in this differential photometry arise from the usual Poisson errors of the count rates inside the photometry apertures for the target and comparison stars, as well as inside the annulus used to determine the sky background.  The sky background is usually small and measured with a large annulus, so this makes for a negligibly small uncertainty in the differential photometry.  For a count of $N$ photo-electrons inside a photometry aperture, the Poisson error will be $N^{-0.5}$.  In practice, we always choose exposure times so that $N \gg 10^4$, so the statistical uncertainties are always $\ll$0.01 mag.  In this case, the statistical errors are always negligibly small.  

CCD photometry programs usually report only these Poisson error bars, so it can be easy to be misled into thinking that the photometric accuracy is greatly better than they really are.  The problem is that there are always ubiquitous systematic uncertainties, usually at the level of 0.01 mag or so.  For as definitive a case as possible, with a different instrumental setup, the one-sigma photometric accuracy in measuring a single optimal V-band magnitude is $\pm$0.0144 mag from Landolt (2009, Table 3), $\pm$0.0069 from Landolt (2013, Table 4), and $\pm$0.0084 mag from Landolt (1992, Table 2).  For a definitive case in measuring the mean errors of a single CCD observation, Clem \& Landolt (2013) give 0.0245$\pm$0.0159 for the V-band.  The Poisson and systematic errors must be added together in quadrature to get the total errors.  With the Poisson errors always being negligibly small for observations of the bright KIC 8462852, the systematic errors always dominate.  The problem is in knowing and minimizing these systematic errors.  For any one observer, looking for a long term light curve, the offsets described in the previous section is not the issue.  Rather, the issue is the image-to-image and night-to-night systematic problems that make for scatter.

The causes of the image-to-image systematic variations are not well known.  Here are some possibilities:  (1) One inevitable effect is ordinary atmospheric scintillation, which might be non-negligible for our typically short exposures of 14-60 seconds.  (2) A related and inevitable effect is variations in the average star profile (as quantified by the PSF's FWHM) across the field as arising due to scintillation, so that a fixed photometry aperture will record differing fractions of the starlight for the target and comparison stars.  (3) The centering of the photometric aperture around each star image will not be perfect, so the fraction of the total starlight inside the aperture will change, and the derived magnitude will therefore jitter at the same level.  (4) Another possibility is small-scale differences in atmospheric extinction across the field, perhaps arising from small clouds or small cells of haze.  (5) Small imperfections and changes in the flat fielding are a ubiquitous and certain cause of photometric variations.  In particular, our flat fielding image series shows variations at the one-percent level on time scales of an hour.  This is particularly pernicious because it means that the real flat field that should be applied to each image is slightly different from the flat field acquired at the start or end of the night, so if the target star is on a slightly low (or high) position in the flat field, then the target will be calculated to be slightly dimmer (or brighter) than a nearby comparison star.  So flat fields have difficulties getting better than 1\% accuracy.  (6) Further, dome flats are never exactly flat illumination of the chip, And in practice, it is not possible to remove all starlight (in particular the outer tails of the star images) from a flat constructed of dark sky or twilight images when the highest accuracy is needed.  (7) A further insidious and unappreciated problem is that the structure in the flat fields is a sensitive function of the color of the incoming light, while incoming star light will never have the same effective color temperature as was used to make the flat field, and the target and comparison stars will always have a somewhat different color and land on pixels with different color behaviours.  So it is inevitable that the color dependence of the flat fields will combine with the various colors of the target and comparison stars to give variations in the measured differential magnitude.  

The flat fielding problem is actually worse than just presented.  (8) The problem is that the structures on the CCD chip make for uneven quantum efficiency across each and every pixel.  These structures include the electronics within each pixel, as well as the microlenses found in all modern CCDs.  That is, flat fields are substantially variable across each pixel, with starlight falling on one part of the pixel being recorded at a higher or lower efficiency as for a nearby region of the same pixel.  So if the peak of the target star image happens to fall on a more sensitive part of a pixel, while the peak of the comparison star image happens to fall on a less sensitive part of a pixel, then the target star will be recorded as being brighter than it should be for the differential photometry.  This can often be a large problem, for example, the {\it K2} follow-on mission of the {\it Kepler} spacecraft finds typically 1\% variations (and even up to over 10\% variations) as the spacecraft suffers drifts at the sub-pixel level (Van Cleve et al. 2016).

In summary, the measured systematic effects even for the best CCD images are of order 0.01 mag, with various effects contributing to this dominant photometric error.  These $\sim$0.01 mag star-to-star and image-to-image systematic changes are not widely recognized, yet they are ubiquitous and dominate the real photometric uncertainties.  

How can these inevitable photometric errors be minimized?  Here are four methods to minimize the systematic errors.
	
(1) To minimize the Poisson errors for the comparison stars, an ensemble of many comparison stars can be used.  This effectively increases $N_{comp}$ so as to make the statistical error bars yet smaller.  More importantly, the systematic errors for the comparison stars will be averaged over many stars, leading to a much more stable standard for differential photometry.  At best, this can only reduce the systematic uncertainty by a factor of 1.4 improvement, because the target star will still have its own systematic error.  

(2) The star images can be placed intentionally somewhat out-of-focus, with longer exposures, and large photometric apertures.  This is no problem for KIC 8462852 because its field is not crowded and it is easy to compensate with somewhat longer exposures.  The extra spreading of the starlight makes for a more even illumination of each pixel (so that sub-pixel variations in the flat field are averaged over) and the inclusion of more pixels with starlight (so the pixel-to-pixel systematic effects are averaged over).  The longer exposures help minimize scintillation variations.  The large photometric apertures means that the excluded starlight is very small, so variations in the fraction of light outside the aperture will be miniscule.

(3) A third method is to place the target star and the comparison stars at exactly the same position on the CCD chip for all images included in the light curve.  This tactic requires an accurate auto-guider as well as consistent practices for years.  Then, any systematic errors that are constant with position on the chip will remain constant, resulting in some small constant offset for the light curve to get to a standard photometric system.  Such errors include those arising from intra-pixel variations in efficiency as well as imperfect flatness of the flat-fields due to the observatory's procedure in exposing the white spot.  This is the tactic used by the {\it Kepler} spacecraft, and is required to give its awesome photometric accuracy.  Indeed, with the imperfect positional stability of the {\it K2} mission, even 0.1-pixel shifts lead to substantial changes, and small drifts in the pointing lead to typically 1\% variations.  This tactic will not recover from systematic effects that change image-to-image or night-to-night, including the observed small changes in the flat field over time scales of hours, scintillation effects for short exposures, small-scale extinction variations, and small variations in the color term due to differences in the airmass.

(4) A fourth method to minimize the systematic errors is simply to average over many individual images.  This averaging can be done by taking many CCD frames in one night, or by taking data on many nights, or by combining observations from many observers.  With multiple observations in one night, all the image-to-image systematics will be averaged out.  With observations on many nights, the image-to-image and night-to-night systematics will be averaged over.  With observations from many observers, their systematic errors will all be completely independent, so all the systematic errors will be beaten down by a factor of the square-root of the number of observers.  

The fourth method is the solution used by Landolt (1992; 2009; 2013) and Clem \& Landolt (2013) for calibrating their standards.  The Las Cumbres Observatory light curve (Boyajian et al. 2018) uses methods 1, 2, and 4 simultaneously to achieve their high-accuracy high-time-resolution light curve.  The {\it Kepler} mission uses method 3 to achieve its incredible accuracy.  In this paper, we will be variously using all of these methods for constructing our light curves.

We are going over the details of systematic errors because they are not widely appreciated, while an easy and wrong view is just to take the Poisson errors at face value.  Indeed, until KIC 8462852, there have been few astronomical photometry programs requiring millimag accuracy from night-to-night for years on end.  The closest example we can think of is with the {\it Kepler} spacecraft, where extraordinary efforts and costs were made to achieve the stability, and yet where very small changes in pointing result in up to 10\% errors in the {\it K2} photometry.  The point is that ordinary ground-based photometry usually has real photometric errors of order 0.01 mag or more, and it requires extraordinary efforts to get millimag photometry that is consistent night-to-night for years.

For seeking secular trends in KIC 8462852 over a few year interval, with known decline rates varying from 0.0011 to 0.035 mag per year, we must measure the light curve to an accuracy of a few millimags or better.  To achieve this, we have variously used a number of the tactics above.

With our individual magnitudes for each CCD image, our first step is to throw out outliers, magnitudes that are more than 5-sigma deviations from the light curve for an individual observer over a nearby time interval.  These outliers are due to the normal problems of hot pixels, cosmic rays, and such, constituting about 1\% of our data.  Pointedly, if these outliers are not rejected, their effect on our final light curve is negligibly small.  Our next step is to form nightly averages for each observer for each filter.  The uncertainty is taken to be the RMS scatter of the nightly observations divided by the square root of the number of input observations.  When many magnitudes are averaged within one night, the quoted error bar can get unrealistically small, with this not including all the systematic errors.  Still, these nightly averages can greatly reduce the Poisson measurement errors and can greatly reduce some of the systematic errors.  The real uncertainty in these nightly averages is then dominated by the night-to-night and observer-to-observer systematic errors.  We have our basic nightly-averaged light curve from all observers together, as tabulated in Table 2.  The columns are the band (B, V, R, or I), the average Julian Date for that night's magnitudes expressed only to the nearest day, the observer, the number of individual CCD frames going into each nightly average, and the nightly averaged magnitude with one-sigma error bar.  This basic data set is based on 19,176 individual CCD images, where the magnitudes are averaged together to form 1,866 nightly averages.  Only the first five and last five lines of Table 2 are displayed in the printed version of this paper (to illustrate format and content), while the full 1,866 lines appear only in the on-line version of this paper.

For each observer and for each filter, we then add a constant offset so as to place all nightly averages onto a consistent magnitude system.  Each offset is determined by minimizing the RMS scatter in the combined light curves for all observers, operating on 20-day bins.  The offsets in B, V, and I for Oksanen, as well as the offset in R for AstroLAB are set to zero, and this effectively sets our standard system.  These offsets are tabulated in Table 1.  

These nightly averaged light curves still have substantial scatter, all due to the expected and normal systematic errors, so we beat down these errors by averaging over many nights and all observers.  In particular, we bin together all nightly averages over successive 20 day intervals.  We chose 20 day intervals because 20 is a round number that is a good balance between good time resolution of the light curve features and having many points in each making for a smaller photometric error.  The binned magnitude is simply the average of the input nights, which implicitly is making the assumption that all the nightly averages have comparable total errors.  This is reasonable as judged by the scatter in the light curves for individual observers.  The calculated 1-sigma uncertainty is again the RMS scatter within each bin divided by the square root of the number of measures.   With this averaging over many nights for all observers, we have substantially reduced the night-to-night and observer-to-observer systematic errors.

Within these 20-day intervals, with the applicable offsets, the average RMS scatters in the light curves are 0.006, 0.004, 0.005, and 0.004 mags for the four BVRI bands.  Within each bin, on average, we have a dozen included nightly averages, so our formal error bars are usually a few millimags, with these being our remaining systematic errors.

With this, we have produced B-, V-, R-, and I-band light curves from 2015.75 to 2018.18, with 20 day time resolution, with these tabulated in Table 3.  The four columns are the band (B, V, R, or I), the average Julian Date of the input nightly averages, the number of nightly averages going into each line, and the average of the input nightly averages along with the one-sigma error bar.  Our four-color light curves are displayed in Figure 1.  Our light curves show considerable structure, and systematic changes with respect to color.  

\section{KIC 8462852 Light Curves}

Overall, the light curve shows a systematic fading.  The best fit line for the B-band has a slope of 0.99$\pm$0.09\% per year, while the best fit line for the I-band has a slope of 0.37$\pm$0.09\% per year.  (These best fit lines were calculated by the usual weighted linear regression.)  The decline from our earliest to latest times (2015.75 to 2018.18) are 0.023$\pm$0.003 mag in the B-band and 0.008$\pm$0.003 mag in the I-band.  These rates of decline are typical of the previously reported secular declines (see Section 1).  This demonstrates that the secular decline from 1890 until the end of the {\it Kepler} run is still continuing even until the middle of February 2018.

However, we see that the secular decline is neither monotonic nor steady.  Our light curve shows three peaks and three dips, all superposed on the general secular decline.  The durations of the three dips are $\sim$120, $\sim$120, and $\sim$180 days.  And indeed, the third dip has the entire Elsie group of dips (Boyajian et al. 2018) superposed on the dip.  This shows that the {\it Elsie} group of dips is just the fine structure superposed on the bottom of a longer dip of duration 180 days.  

For comparison, we can look for month-long dips in the {\it Kepler} full frame images (Montet \& Simon 2016), which cover nearly four and a half years.  The {\it Kepler} light curve is not composed of many shallow 120--180 day dips.  But it does display the first half of a 600 day 3\% dip, with a superposed series of $\sim$1\% dips lasting around 80 days (from {\it Kepler} day 1490--1570), with further $\sim$20\% day-long dips superposed at the bottom of the broader dips (Boyajian et al. 2016).  (We can speculate that the other dips also contain some number of short-duration dips, all of which add together to make the overall dip with durations of 120--180 days.)  This also shows that the secular decline has many superposed short-duration dips.

From Figure 1, we see that the B and V light curves share nearly identical structure, while the R and I light curves also share similar structure.  Indeed, when the curves from Figure 1 are all shifted vertically to have the same average, we see that all of the B, V, R, and I light curves share the identical structure.  These different filter light curves have most systematic errors that are completely independent.  So the identical structures provide strong evidence that the structures are not caused by systematic artifacts.  Further convincing evidence comes from the identical structures appearing in the completely-independent light curves of our many observers.  A statement and proof to this point is not usually needed, but in the case of KIC 8462852 in 2.43 years, the light curve structure has amplitudes of order 0.01 mag, so everyone should worry about systematic effects.  A lesson from this is that small effects such as here in this paper must have something like many observers and multiple colors to prove against systematic problems, or else have extraordinary methodology (such as for the {\it Kepler} spacecraft). 

We have quantified the shape of the light curve structures by measuring the slopes and amplitudes over each of five different sets of time intervals.  The slopes were calculated from the magnitudes in Table 3 with a chi-square fit to a straight line over the time intervals given in Table 4.  The amplitudes were calculated from the differences in the weighted averages for the time intervals in Table 4.  Table 4 further lists these slopes and amplitudes for each of the BVRI light curves.

Table 4 shows that the bluer colors systematically have higher slopes and higher amplitudes than the redder colors.  This can be quantified as the ratios of the slopes and amplitudes for each color relative to the B-band values.  Each ratio has typically 11\% to 18\% uncertainty.  All the ratios are consistent within each color.  Further, this chromatic extinction is applicable to both the overall secular dimming (from either the overall slope or the amplitude from start to end) and the 120--180 day dips (from the slope and amplitudes around the {\it Elsie} group).  This means that the light curves for each color have similar shape, except for a scale factor compressing or extending the vertical dimension.

So we now have the relative variation for several sets of time intervals in the V-, R-, and I-bands with respect to the variations in the B-band.  From each line in Table 4 (with each line representing a different set of time intervals), we can divide the amplitudes/slopes for each color by the values for the B-band.  These ratios should equal $A_V/A_B$, $A_R/A_B$, and $A_I/A_B$ for dust extinction.  We have five measures of each of these extinction ratios.  By averaging the extinction ratios for each of the five lines, we can get average extinction ratios in each color with 11\% to 18\% error bars.  The first line in Table 4 is not completely independent from the next four lines due to relatively small overlaps in the magnitudes included in each fit.  These averages are $A_V/A_B$=0.77$\pm$0.08, $A_R/A_B$=0.66$\pm$0.12, and $A_I/A_B$=0.36$\pm$0.06.  

A more general way to get the extinction ratios is to look at the slope of the plot of V versus B, and so on.  With the usual linear fits, we get $A_V/A_B$=0.77$\pm$0.05, $A_R/A_B$=0.50$\pm$0.05, and $A_I/A_B$=0.31$\pm$0.05.  This is within error bars of our prior result.  We take this more general result to be our best measure of the chromatic extinction.

These slopes, amplitudes, and ratios are simply descriptions of our light curve, where we have not specified what is a dip and what is a secular dimming.  A light curve that shows some apparent secular fading could well be just displaying the ingress of some long-duration dip.  The obvious description of our light curve is a secular dimming with three 120--180 day dips superposed.  In this case, we have measures of the chromaticity of the extinction for the secular dimming (lines 1 and 5 in Table 4) and for the dips (lines 2--4 of Table 4), and we see that both the dipping and dimming are achromatic with similar color effects.  Alternatively, our light curve could be said to have no secular evolution, but the three dips superpose in such a way to mimic a secular change over our 2.43 years.  In this case, we have no measure of the chromaticity of the secular evolution.  (But Davenport et al. 2018 and Meng et al. 2018 already show that the secular evolution has a chromaticity comparable to our findings.)  We are not able to say whether the general decline from 1890--1990 is just the ingress of a very-long dip, nor whether the 120--180 day duration minima in our light curve are just the fast component of the secular decline. 

What we see in Figure 2 is variations on all time scales from hours to a century, such that a power-density-spectrum of the light curve would show significant power over a very broad and continuous range of frequencies.  (The highly non-uniform light curve sampling, residual uncertainties in the offsets between the light curve segments, and the extreme variations in light curve density all conspire to make it impossible to construct an actual power-density-spectrum of any useful reliability.)  From Figures 2 and 3, KIC 8462852 displays dips with a very wide range of dip durations.  Specifically, we see may dips with durations 0.4--10 days with {\it Kepler}, the Elsie group of dips have durations of 0.3--14 days, our light curve shows dips with duration 120--180 day durations, our joint light curve shows a 2-year duration dip centered on 2013.0, the Harvard light curve shows a 10-year duration dip from around 1900--1910, our joint light curve shows a $\sim$56-year dip from 1950--2006, and the whole light curve from 1890 to 2018 looks like 60\% of a dip that would have a duration of perhaps 200 years.  All throughout, we see dips superposed on dips superposed on dips.  There does not appear to be any bimodality in the duration of the dips, which is to say that all dip durations from a century down to a day are populated with observed dips, so the duration distribution forms a continuum with no substantial gaps at any intermediate time scale.  This is making the point that we have no empirical means of making a useful dividing line between dipping and dimming.

Our `nightly averaged' light curve has had observations from the individual observers averaged over a single session at the telescope, typically a few hours or less in duration, so the many points in Table 2 actually have a time resolution of a few hours.  We have analyzed this fast light curve for effects that would be averaged over with 20-day binning.  We do not see any significant periodicity at the 0.88 day apparent rotation period (Boyajian et al. 2016).  We did not expect to see any such photometric period, because its amplitude is well below the 0.001 mag level.  Further, our light curve does detect the four dips from the Elsie group, but our dips are not highly significant, mainly because our time resolution and time coverage during the dips are not adequate to resolve the dips.  For another test, we have examined our time series photometry for variations even down to 40-second in duration, but we find no significant dips or flares on any fast time scales.

\section{The Light Curve from 1890 to 2018}

Our light curves are just part of an array of light curves covering the last decade and century.  We can get a good perspective on the overall evolution of KIC 8462852 from 1890 to 2018.18 by plotting all these light curves together.  Each of these was taken with different filters and comparison stars, so there is some vertical offset for each light curve to place it on some standard photometric system.  These offsets are difficult to know with any high precision, so we have treated these offsets as free parameters that can be adjusted over some small range, and these offsets were set by minimizing the differences between the light curves.  This procedure of minimizing the scatter in time bins does not work for cross-calibrating light curves separated in time with no one source spanning the gap (like from 1992--2005), and this issue will be corrected as discussed below.  Further, we have just shown that the vertical structure scales in size with the wavelength, so in principle, adjustments need to be made, for example, compressing the B-band light curves vertically and stretching the r'-band vertically.  With these irregularities, we cannot make a perfect overplot of all the light curves, although we are confident that our overplotting with vertical shifts will clearly show the nature of the full evolution.  With this. we have over-plotted the light curves from this paper, from Harvard plates, Maria Mitchell, Kepler, Las Cumbres, ASAS-SN, and from ASAS as displayed in Figure 2.  The area from 2006 to present is crowded and complex, so a blow-up is presented in Figure 3.

A critical issue in the construction of this 1890--2018 light curve is the gap from 1992--2005 where we have no magnitudes.  The problem is that we do not have one observatory that reports magnitudes on both sides of the gap, so there can be some vertical shift between the 1890--1991 and 2006--2018 combined light curve segments.  The shift or offset across this gap is critical for dust models, as this sets how much dust is present for the modern baseline.

A possible solution would be to get the Johnson B-magnitude for KIC 8462852 from the 1890s from the Harvard plates, as well as the Johnson B-magnitude for some recent data set, then the relative positioning across the gap would be determined.  The native system of the Harvard plates is very close to the Johnson B-system (indeed, the Harvard plates were the historical original basis for the Johnson B-system), while the comparison stars (from APASS) are in the Johnson B-system, so the DASCH magnitudes for the Harvard plates are closely in the Johnson B-system.  So the earliest and brightest decade in the Harvard light curve is $B=12.274\pm0.013$ for the time bin 1895$\pm$5 (Schaefer 2016, Table 2).  In principle, we can take the B-magnitude from some recent light curve and place the offset across the gap.  But this solution does not work.  The problem is that the  modern data has a tremendous scatter in the B magnitude, from 12.26 to 12.80 (see Table 4 of Simon et al. 2018).  This huge scatter cannot be due to variability of Boyajian's Star (see Figure 3).  The same large scatter is seen for all the nearby constant stars with similar magnitude and color.  Rather the problem is with the cross calibration of the many sources.  

The various modern sources are all quite accurate for {\it relative} magnitudes, so a good solution is to use differential photometry between KIC 8462852 and some comparison star(s) for both the Harvard data as well as for the modern data.  That is,  the fading of the star between the two epochs is  
\begin{equation}
\begin{split}
\Delta = B_{modern}-B_{Harvard}=(b_{modern}-b_{comp,modern})- \\
(b_{Harvard}-b_{comp,Harvard}),
\end{split}
\end{equation}
where the lower case ``b" indicates the reported magnitudes in the data set, and the data sources and stars are indicated in the subscripts.  As long as stars of closely similar color are used, all the various color terms cancel out, and the resulting differences in magnitudes can be closely compared between data sets.  We have done this with three modern data sets, all coming up with closely consistent values for $\Delta$.

So the task is to compare two baseline time intervals on either side of the gap, so we can cross calibrate the modern and archival light curves.  For the archival baseline, we should choose the 1890s, because this is the time interval when the star was the brightest, and hence this is relevant for knowing the limit on the modern steady extinction.  For the modern baseline, we have chosen the interval JD 2457510 to 2457570 (2016.33 to 2016.50), because this is the well-observed broad maximum in our light curve, and thus has the minimum extinction for our data.  Other baselines are practical for other data sets to quantify the brightness before a dip.  For example, a baseline just before the Elsie dip serves as a `zero' for measuring the dust from the clump that created the Elsie dip.  But this baseline does not provide a measure of the steady dust contribution, nor the total extinction at the time of the infrared observations.  For these various modern baselines, the offsets can be determined with the ASAS light curve that covers the whole time interval.

Our first determination of $\Delta$ makes use of our own B-band data.  For the time interval JD 2457510 to 2457570 (our standard modern baseline), our magnitudes across several observers show that KIC 8462852 is consistently 0.647$\pm$0.001 mag fainter than the comparison star TYC 3162-1001-1 (AAVSO 113).  From the 1895$\pm$5 Harvard data, for this same comparison star, we have $B=11.741\pm 0.013$ (Schaefer 2016, Table 1), so the star KIC 8462852 is 0.534$\pm$0.019 mag fainter.  The difference in these differences shows that the Boyajian star faded a total of $\Delta$=0.113$\pm$0.019 mag from 1895 to 2016.42.

Our second determination of $\Delta$ gets the differential magnitude of KIC 8462852 with respect to four comparison stars (all very close in magnitude and color) as taken from the APASS catalog and the Harvard light curve.  The APASS magnitudes were taken from five visits with BVgri measures between 6 June 2010 and 28 October 2012.  The four comparison stars are TYC 3162-1320-1, 3162-808-1, 3162-1001-1, and 3162-1073-1, with $B$ values of 12.133, 12.382, 11.721, and 12.715, as well as $B-V$ values of 0.543, 0.572, 0.458, and 0.636 mag respectively.  These same stars have Harvard 1890s B magnitudes of 12.151, 12.410, 11.741, and 12.703 respectively.  The Harvard-to-APASS differences in the B magnitudes for the comparison stars averages out to 0.013 mag, which means that the APASS and Harvard magnitudes are closely on the same photometric system.  The RMS scatter is 0.018 mag, for an uncertainty in the ensemble of four stars of $\pm$0.009 mag.  Compared to the average of the four comparison stars, the Boyajian Star was 0.023 mag fainter in the 1890s and was 0.122 mag fainter in 2010-2012.  With this, the Boyajian Star faded by $\Delta$=0.099 mag from the 1890s to 2010--2012.  The errors are correlated across these differences, the errors arising from the comparison stars are relatively small with four stars being averaged together, so the total uncertainty in this offset is dominated by the measurement of the single target star in the 1890s, so the total error will be near $\pm$0.016 mag.  The ASAS data shows that the baseline in 2010--2012 is close to the brightness in our standard baseline, so no offset is needed.

Our third determination uses the B magnitudes from the Las Cumbres photometry (Boyajian et al. 2018) for a thirty day baseline just before the {\it Elsie} dip.  The time range for this baseline is JD 2457850 to JD 2457880.  The useable comparison stars are TYC 3162-808-1 (AAVSO 118) and TYC 3162-1001-1 (AAVSO 113).  Going by the two comparison stars, Boyajian's Star has faded by 0.205$\pm$0.027 and 0.137$\pm$0.019 mag from 1895 to just before Elsie.  With averaging the two measures, the fading is $\Delta$=0.171$\pm$0.016 mag.  Our B-band light curve shows that the target star faded by 0.011$\pm$0.004 mag from the standard baseline until the baseline just before the start of {\it Elsie}.  With this adjustment, we see that KIC 8462852 faded by $\Delta$=0.160$\pm$0.017 mag from the 1890s to the modern baseline.

We now have three determinations of $\Delta$ from the 1890s to the JD 2457510-2457570 baseline; 0.113$\pm$0.019, 0.099$\pm$0.016, and 0.160$\pm$0.017 mag.  The average and weighted average is 0.124 mag.  The uncertainty in this average could be taken as the RMS scatter divided by the square-root of 3 (0.018 mag), the sigma from the weighted average (0.010 mag), or simply as the dominant error from DASCH for the 1890s (0.013 mag).  So, we get a final offset from 1890.0--1900.0 to 2016.33--2016.50 of $\Delta$=0.124$\pm$0.013 mag.

This offset was used in the construction of Figure 2.  We see that the Boyajian Star experienced a rise in brightness by roughly 0.09 mag between 1990 and 2006.  With this, we have what looks like a dip from around 1950 to 2006, with a duration of half-a-century.

With the realization that the dust coverage in the 1890s cannot be below zero, we can now put minimum values for the steady extinction at several critical times.  For the baseline relevant for the dips in the {\it Elsie} group, there was at least 0.135$\pm$0.013 mag of dust extinction in the B-band.  With the ASAS light curve providing the connection, we see that the first very-deep {\it Kepler} dip was superposed on a baseline of near 0.13 mag extinction operating in the B-band, while the later very deep dips were superposed on a time with steady dust coverage with 0.144 mag extinction.  

The most important offset (relative to the minimal extinction of 1890) is for the dates when the infrared limits were made.  The point is that the modern infrared measures must have been made with some amount of steady dimming, and this level is what is needed to place constraints on the total dust around the system.  Wyatt et al. (2017) summarizes the history of all the infrared observations seeking excess thermal emission from KIC 8462852, plus they use all of the observations to individually place a limit on the quantity of dust in their generic dust model.  They find that the most restrictive observations is the 12 micron limit placed by the {\it WISE} spacecraft on 14 May 2010 during its initial cryogenic phase.  This most critical measure was made on JD 2455331, near the start of the {\it Kepler} light curve and in the middle of the ASAS light curve.  The relative offset from 2010.37 to 2016.33--2016.50 is near zero.  So the extinction at the time of the most-restrictive infrared observation is near 0.124 mag more than the 1890 case, all in the B-band.

\section{Implications}

Our light curves have the implication of confirming the existence of the long-term secular dimming, in this case over 2.43 years, but the need for such confirmation has now long passed.  Rather, our light curves have the implication that the secular dimming is going on even {\it now}.  The secular dimming is not just some historical relic from the last century, of interest only to historians.  With the secular dimming on-going, it means the astronomical community can use sophisticated modern techniques on the phenomenon.  Perhaps some aspect of the dimming will be more readily recognized, with application to understanding the fast dips.  Further, as the secular dimming and the {\it Kepler} dips are from the same physical mechanism, then the on-going secular dimming can be studied in place of waiting for short-duration dips.

Our light curves have a sufficient time resolution and time coverage to show the multiple dips structure of the star's secular dimming on the time scale of months and years.  We see that the overall secular decline is superposed with dips with time scales of 120--180 days, plus the {\it Elsie} group of dips with time scales of 1--10 days superposed.  Further, with the 1890--2018 light curve, we have the first view of the complex ups-and-downs, including a 50 year dip (from roughly 1950 to 2000).  The distinction between dipping and dimming becomes blurred.

Our light curve provides the connection between the century-long secular decline and the fast dips.  That is, the  {\it Kepler} and {\it Elsie} dips have time scales ranging from 0.3--14 days, while the archival plate data from Harvard and Maria Mitchell observatories show dips with a time scales of 10 years (1900--1909) and 50 years (1950--2000), plus a century-long overall decline that is likely just some ingress into a very long dip.  Now, our new light curves with dips of durations 120--180 days fills the gap in the time scales.  Further, our joint light curve in Figure 3 shows a 2-year dip centered on 2013.  So now we see durations and timescales of hours, a day, a week, a month, a year, a decade, and a century.  With this it is obvious that the variability time scales in KIC 8462852 form a continuum, all working simultaneously.  As a continuum, then the physical mechanism for the fast {\it Kepler} dips would be the same as the physical mechanism for secular dimming.  That is, all the variability in KIC 8462852 arises from just one cause, and this has some parameter that varies over a wide range, creating dips and dimmings over a wide range of time scales.

Our light curve shows substantial structure in the overall secular dimming, and these structures scale in size only with the wavelength of the light.  That is, the B-, V-, R-, and I-band light curves are identical except for a vertical scale factor.  The secular dimming is 0.77$\pm$0.05, 0.50$\pm$0.05, and 0.31$\pm$0.05 times that in the blue, for the V-, R-, and I-bands.  (These measures apply only to our 2.43 years of light curve, including the {\it Elsie} groups of dips, while the chromaticity could conceivably vary with the time scale, amplitude, or year.)  These factors prove that the secular dimming (merging into the short dips) are not caused by occultations of the primary star by any object of high optical depth.  Occultations by planets or stars or solid bodies or any opaque cloud will cause achromatic secular dimming.  This allows us to rule out a wide array of possibilities for mechanisms of the secular decline.  However, the observed chromatic fading is consistent with the prominent hypothesis that the occulter is a dust cloud.  This good agreement with the dust-extinction model provides strong physical evidence that the secular dimming is caused by the occultation of the central star by a dust cloud that changes its opacity in front of the star.  That is, apparently the secular dimming is caused by dust clouds passing in front of the star.

The three extinction ratios can be compared to the ratios predicted for canonical ISM dust extinction (with R=3.1) as 0.76, 0.57, and 0.37 (Mathis 2000).  For typical circumstellar dust, an R value of 5 is more appropriate, with ratios of 0.83, 0.66, and 0.46 in B, V, and R respectively (Mathis 2000).  Our measured ratios are within reasonable agreement, to within the error bars for both R=3.1 and R=5, although the agreement is better for R=3.1.  This can be compared to independent estimates of $R \gtrsim 5$ from Meng et al. (2018), $R\sim5$ based on $B/i'=1.94\pm0.06$ from Boyajian et al. (2018), and $R=5.0\pm0.9$ from Davenport et al. (2018).

This chromatic extinction implies that the dust  must have small particles, comparable in size to the wavelength of optical light.  From Figure 3 of Wyatt et al. (2018), we see that dust distributions with a minimum size of 2.3 microns results in achromatic extinction in optical bands, while a minimum size of 0.1 micron results in color effects comparable to that observed in the dipping/dimming.  

A feature of this small dust is that it will be rapidly accelerated by the ordinary stellar radiation, and any freshly created cloud of dust will be rapidly dispersed.  The ratio of radiation forces to gravitational forces is independent of distance from the star, but the radiation pressure will become relatively large for small particles.  This radiation force will be in the radial direction, but ordinary orbital mechanics will rapidly spread out the dust tangentially in the plane of their orbit, and this dispersion will greatly decrease the dust optical depth along the line of sight to the star.  This puts some upper age limit on the dust clouds, with the limit depending on the details of the dust.  The blowout time near periastron is about the same as the time the dust spends around periastron, as long as the grains are not too small.  For reasonable orbital parameters, the blowout time will be a small number of months.  That is, we cannot have the observed dust be simply in some stable orbit around the star for many orbits.  Rather, the dust cloud creating both the dipping and the dimming must have been made recently, presumably as something approaches the star.  The dipping and dimming has been going on for a century, so we must have many bodies repeatedly approaching the star for the last century, with each one making a dust cloud where the small particles are rapidly dispersed.  In this scenario, the secular dimming would come from some secular increase in the number or size of bodies passing through periastron.

We now have three strong arguments that the secular dimming and the fast dips are caused by the same physical mechanism:  (1) The continuum of time scales for the variability, from hours to a century, is easily explained as the variability of a single parameter within a one-mechanism model.  Whereas, a two-mechanism model would require special pleading to make the continuum as observed.  (2) Both the secular dimming and the fast dips are chromatic in a way as predicted by the dust cloud extinction mechanism, and this rejects many other mechanisms with the occulter being a solid body or opaque cloud.  So, there is a good case for both dips and dimming to arise simply from occultations by dust clouds.  That is, dips and dimmings are caused by the same mechanism -- dust clouds.  (3) The original connection was made by Schaefer (2016), as based on an Ockham's Razor argument.  Both the secular dimmings and the fast dips are completely unique and inexplicable phenomena for an isolated middle-aged F3 main sequence star, with no non-photometric peculiarities yet seen.  These very particular and unique properties both require a rare mechanism, and either one or two mechanisms are needed.  Ockham's Razor decries the multiplicity of hypotheses for explaining a set of phenomena, so we have a strong point that both dips and dimmings result from one cause.  Taken together, these three arguments come close to a proof that only one physical mechanism (i.e., occultation by dust clouds) is involved.

A critical implication is that the recent average dimming is 12.4\% over the historical minimum.  This historic baseline is set by the brightest part of the light curve (in the 1890s) for which the dust coverage will be at a minimum.  This implies that a lot of dust extinction is a part of the modern baseline.  We can make an order-of-magnitude estimate of the total dust needed for the last century of coverage.  The dust covering at any time rapidly passes across the star's front.  While the passage time across the star varies substantially depending on the presumed orbit of the dust, the typical passage time corresponds to something like the shortest observed duration of around one day.  The deepest {\it Kepler} dip was about one day in duration, representing one small relatively dense cloud, and we can take the total dust in this cloud as a unit amount of dust.  So a dimming of around 20\% lasting for a day will need a total dust equal to this unit.  The dust that passes along the line of sight will be proportional to the duration and the depth, so for example a 10\% drop lasting for ten days would require around 5 units of dust.  From Figure 2, we see that the dimming has been going on for more than a century (36500 days).  Our result is that the dimming is by $\geq$12.4\% inn depth at the end.  But the dimming has been roughly linear over the century, so the required dust is only half that  needed if the entire century had a 12.4\% drop.  With more than a century of secular dimming, we need something like $(36500/2)*(12.4/20)=11,300$ times the dust as from a single deep Kepler dip of duration one-day.  To round numbers, all that is justified by the order-of-magnitude calculation, the secular dimming over the last century requires $10^4\times$ as much dust as is needed for the Kepler dips.  

The total mass in this dust can be estimated from the result of Bodman \& Quillen (2016), with the deepest {\it Kepler} dip requiring just under $10^{-7}$ Earth masses in the dust alone.  So the secular dimming then requires $\sim10^{-3}$ Earth masses of dust.  This does not count dust out of the plane, nor any additional steady dust for the coverage in 1890.  Further, this does not count non-dust material that might come along with the dust.

This dust must be spread around the star system, covering a substantial part of the sky.  The chromatic extinction for both the dipping and the dimming implies that the dust cannot be hidden in optically thick regions.  And with such broad coverage of dust in one plane, the coverage perpendicular to the plane might be substantial, so covering yet more of the sky around the star and requiring more dust.  Further, if there is substantial dust coverage in 1890, then the modern dust coverage will be substantially larger.  In all, the amount of dust covering the sky around KIC 8462852 must be $\gg$10,000 times as much as is required for a single 20\% {\it Kepler} dip of duration one-day.  To get the dust for a single {\it Kepler} dip, we are already forced into extravagant scenarios, for example involving tight swarms of super-comets.  And now the problem to get all the required dust is greatly larger.

This 12.4\%-low modern baseline (as compared to the historic baseline with the minimal dust coverage) has the implication that the modern infrared observations were made during a time when there was large amounts of warm dust.  This sets up problems with all models for the dust.  Wyatt et al. (2018) provide exhaustive and detailed calculations of a wide range of generic dust models and their infrared emissions.  In general, the occulting dust clouds must be on some elliptical orbit, and making a high eccentricity allows for the dust spending most of the time far from the star with little near infrared emission.  The observable IR light is dominated by the dust during their periastron passage.  Given the short duration of the {\it Kepler} dips (0.4 days being the fastest), the occultations must occur around the time of periastron.  So the current dust coverage is a direct measure of the amount of hot dust around periastron.  If we had no Harvard light curve, then we'd be free to expect that the modern baseline has near-zero dust covering, and hence the infrared light should be so faint as to result in no measurable infrared excess.  But this is not the case, as the Harvard light curve shows the real baseline with minimal extinction is 12.4\%.  So that means that there must be a long string of dust clouds, spread along a Keplerian orbit, passing around periastron, emitting substantial amounts of infrared emission exactly at the times when the IR searches were being made.

Wyatt et al. (2018) provide detailed calculations of the IR excess for two cases, where there are 0\% and 16\% dust coverage (in the {\it Kepler} band) at the start of the {\it Kepler} run.  The 16\% case is strongly rejected, because it predicts that both {\it Spitzer} and {\it ALLWISE} measures should have seen a strong IR excess on many occasions.  The model violations of the many upper limits varies from 3$\times$ to 10$\times$ (Wyatt et al. 2018, Figure 8).  A more general analysis comes from Figure 6 of Wyatt et al. (2018), where we can see the effects of changing the orbital periastron distance, the orbital eccentricity, the dust type, and the orientation of the periastron with respect to the line of sight.  This figure shows a curve that limits the star-cloud distance such that it is possible to get a dip as short as 0.4 days in duration.  Their figure also shows a curve for the case where the baseline dimming (for application to the times of the infrared limits) is 10\% for the Kepler band-pass.   Their Figure 6 defines the allowed range of orbital and dust parameters for the general dust model consisting of dust clouds strung out along a highly eccentric orbit.

For comparison with our limit on the smooth dust coverage at the time of the most restrictive infrared observation ($A_B = 0.124 \pm 0.013$ mag), we cannot directly use the Figure 6 of Wyatt et al. (2018).  One difference is that our limit is in the B-band, while their Figure 6 is for the steady extinction coverage as seen in the {\it Kepler} band.  The second primary difference is that the dust used for their Figure 6 has an adopted minimum dust size of 2.3 microns, chosen so that the dust will not be rapidly blown away and dispersed by the starlight, but this makes for achromatic extinction (see Figure 3 of Wyatt et al. 2018).  The dust we are seeing has a fairly strong dependence of extinction on wavelength, so there must be smaller dust particles.  The small grains have emission efficiency at 12 micron that will be lower relative to the extinction at much shorter wavelengths, thus allowing for a higher level of dimming. This is counteracted to some extent by the increase in temperature of the small dust grains caused by its lower emission efficiency, but not by as much, so that the overall effect is for the non-detection of thermal emission to provide less stringent constraints on the level of dimming allowed as compared to Figure 6 of Wyatt et al. (2018).

So we have repeated the model calculations except where the steady extinction depth ($max(\delta_B$, the largest fractional loss of light) is calculated for the B-band, and the dust distribution is the usual power law extending down to 0.1 micron particle size.  With these two changes, we have constructed our Figure 4, with the same format and input as as described in Figure 6 of Wyatt et al. (2018).  This will quantify whether our limit on the steady dust coverage is compatible with the general dust model.

In our Figure 4, our limit on the extinction depth forces the case to be close to the labeled `0.1' curve, or somewhere to the right of that curve.  At the same time, fast duration of the {\it Kepler} dips forces the orbit to be to the left of the dashed curve labeled `0.4d'.  Further, the lack of any periodicity in the {\it Kepler} light curve forces the orbit to have parameters lying above the diagonal black line.  So we have a large region of orbital parameters, roughly from 0.05--0.5 AU for the periastron distance, which is compatible with our steady extinction, the minimum dip duration, and the lack of periodicity.  The critical curves shift together as the longitude of the occultation changes, with an allowed range going from 0.01--0.1 AU for a longitude of 90$\degr$ occultation.  This detailed set of calculations for the default model shows that we have a simple and consistent model that can explain the physical setting of the dips and dimming.  One caveat is that the dust required to reproduce the colour dependence of the extinction is small enough that it should be removed by radiation pressure and so not be distributed around any elliptical orbit. The physical interpretation of the elliptical orbit in the model could instead refer to that of the larger dust grains and fragments created in the break-up of a comet, with the small dust causing the extinction created as these bodies pass through periastron (with our line of sight oriented just after this).

Two other published measures give the {\it total} extinction in modern times.  The first is from Boyajian et al. (2015), where the extinction was measured to be $E(B-V)$= 0.11$\pm$0.03, implying $A_B = 0.66 \pm 0.24$ mag for $R=5$ dust.  The second measure is from Meng et al. (2018), where they have fitted stellar models to the spectral energy distribution and report that $A_V = 0.73 \pm 0.05$.  If this dust is orbiting the star, as in the generic dust model, then the lack-of-infrared-excess constraint greatly violates the limit forced by the short durations of the fast {\it Kepler} and {\it Elsie} dips (see Figure 4).

It is possible that this total extinction arises from incidental dust in the intervening interstellar medium.    Indeed, this might even be the expected case, for a canonical average visual dimming by 1.8 magnitude per kiloparsec in the plane of our Milky Way and KIC 8462852 being roughly 0.45 kiloparsecs distant.  In this case, the best and only constraint we have on the extinction from dust near the star at the time of the infrared observations is our new limit that $A_B \geq 0.124 \pm 0.013$ mag.    However, for a measure specific to KIC 8462852, as based on the sodium line depth, Wright \& Sigurdsson (2016) ``require that Boyajian's Star be suffering roughly 35\% extinction [in the V-band] due to interstellar dust, and not significantly less or more."  With this, for the {\it total} extinction of Meng et al. (2018) with $A_V = 0.73 \pm 0.05$, we have roughly 0.38 mag of extinction from circumstellar dust.  This would be inclusive of the $A_B = 0.124$ mag difference we found from the 1890s to recent years.  As such, we would attribute the difference to a steady dust component that was already present in the 1890s.  With this extinction of  $A_V = 0.38$ for the circumstellar dust alone, we then find that the general dust model is then confined to a small region of parameter space.  That is, in Figure 4, the limit curve corresponding to $A_V = 0.38$ lies right on top of the dashed limit curve from the minimal duration.  The position of this allowed region can be moved around by changing the longitude of the occulting dust, while the size of the allowed region can be greatly enlarged by allowing dust smaller than 0.1 micron in size.  Alternatively, it is easy to think that the interstellar extinction is substantially larger than found from the sodium absorption line depths.  So, in all, the general dust model with ordinary dust can be made to agree with all the data.

So, we are finding that our limit ($A_B= 0.124 \pm 0.013$ mag at the time of the most constraining infrared observation) is easily compatible with the general dust model.  However, there are two cases where this easy compatibility goes away.  The first case is if there is substantial dust coverage in the 1890s.  We have no reason to think that the earliest observations just happen to be the time with near-zero extinction.  Indeed, with the reported difference between total extinction and interstellar extinction being $A_V = 0.38$ mag, the circumstellar extinction is apparently greatly larger than our limit.  The second case is where there is substantial amounts of dust above or below the orbital plane that does not occult the star.  Such dust would not affect the durations and amplitudes of the dipping and dimming, while the extra dust might increase the infrared excess to the point where it violates the observational limits.  With our limit, having a mere twice as much dust above/below the orbits that occult the star will result in three times the infrared excess and then violate the constraints in Figure 4.  Both of these two cases are plausible, and perhaps likely.

The interpretation of the dipping/dimming as arising from circumstellar dust is now in many ways inevitable, but our community is still faced with the unanswered dilemma of how to get the dust to explain the dips.  This is expressed in the original title question of ``Where's the flux?" (Boyajian et al. 2016).  Now, the limit from our paper is compatible with the general circumstellar dust model with the dust generators strung out along a highly eccentric orbit, creating dust as it approaches the star, soon to be dispersed.  Still, with substantial circumstellar dust coverage in the 1890s or off-plane dust, the general trouble with all models involving dust is the lack of any detectable infrared excess. This gives a second meaning to the question ``Where's the flux?".

\section*{Acknowledgements}

We are deeply grateful for the many people who contributed funding for the Kickstarter campaign "The Most Mysterious Star in the Galaxy", as this provided the support for the Las Cumbres Observatory data that is so perfect for the light curves of the {\it Elsie} group, as well as for our calibrations.  The {\it American Association of Variable Star Observers} ({\it AAVSO}) has provided much that was required for all of the the observers for our program, including finder charts, comparison star magnitudes (through the {\it APASS} program), programatic advice, and a data archive.  Funding for {\it APASS} has been provided by the Robert Martin Ayers Sciences Fund.  The DASCH data from the Harvard archival plates was partially supported from NSF grants AST-0407380, AST-0909073, and AST-1313370.



\begin{figure*}
	\includegraphics[width=\textwidth]{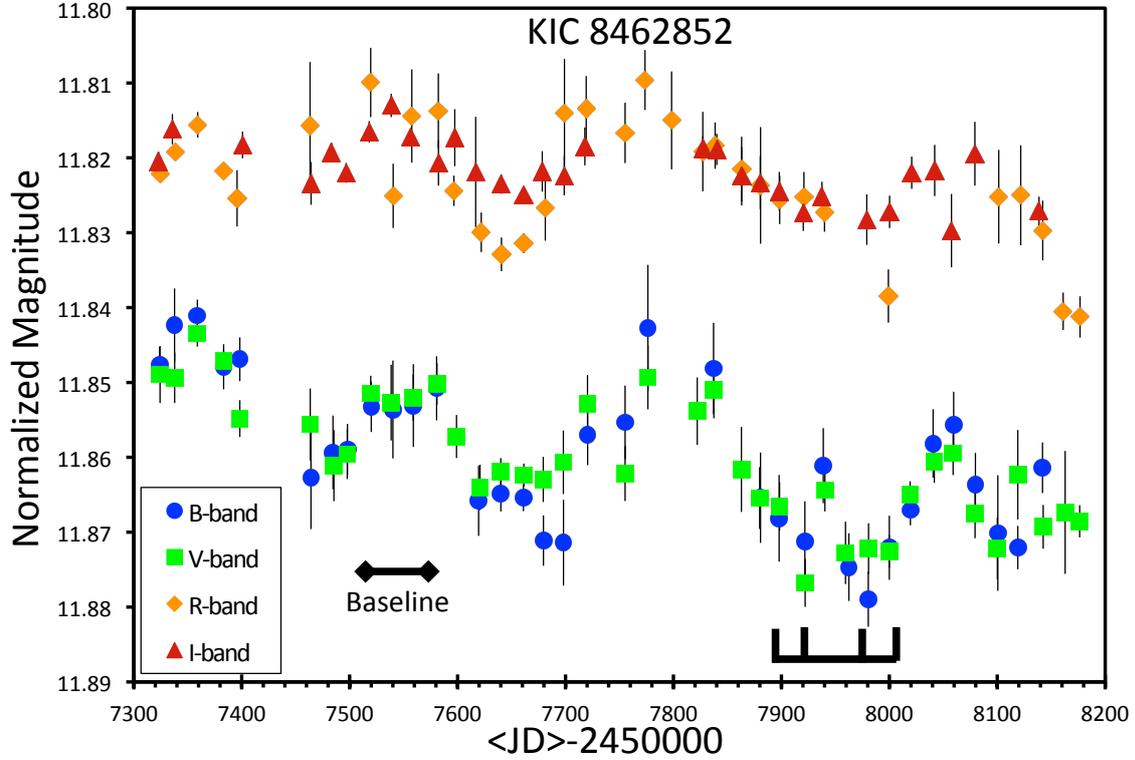}
    \caption{BVRI light curves of KIC 8462852 from 2015.75 to 2018.18 averaged into 20-day bins.  The light curves have been shifted vertically so that the B overlaps the V, and the R overlaps the I.  The V-band light curve is unshifted.  In the bottom half, the B points are blue circles and the V points are in green squares, while in the upper half, the R points are orange diamonds and the I points are burnt-red triangles.  For illustration purposes, the points tabulated in Table 3 with quoted error bars of 0.009 mag or larger are not plotted.  The horizontal bar with four upward ticks, near the bottom right, indicates the times of the {\it Elsie} group dips, {\it Elsie}, {\it Celeste}, {\it Skara Brae}, and {\it Angkor}, from left to right.  With our 20-day binning, we do not resolve the {\it Elsie} complex of dips, but the binned light curve is at its minimum during the dips.  We have selected the time interval JD 2457510 to 2457570 (2016.33 to 2016.50) as our modern baseline because it is the well-measured broad peak of our light curve, with this time interval represented by the horizontal bar just to the right of the legend.  One point that we see from this figure is that the detailed variations in B are closely matched by those in V, as well as the detailed variations in R are closely matched by those in I.  The critical point of these light curves, and the reason for our observing program, is that we see a secular decline over the 2.43 year interval.  In the B-band, the decline from the start to the end is 0.023 mag, or 1.0\% per year dimming.  This secular dimming is at a similar rate as has been seen by many spaceborne and ground-based programs ever since 1890.  However, this secular dimming is certainly not monotonic, and we see three apparent peaks and dips.  The dips have durations from roughly 120 to 180 days, with the {\it Elsie} group of short dips superposed.  A further critical point to be seen from this figure is that the bluer light curves have larger amplitude variations and steeper declines than the redder light curves.  This behaviour is as predicted for the secular dimming being caused by dust clouds occulting the parent star,}
\end{figure*}

\begin{figure*}
	\includegraphics[width=\textwidth]{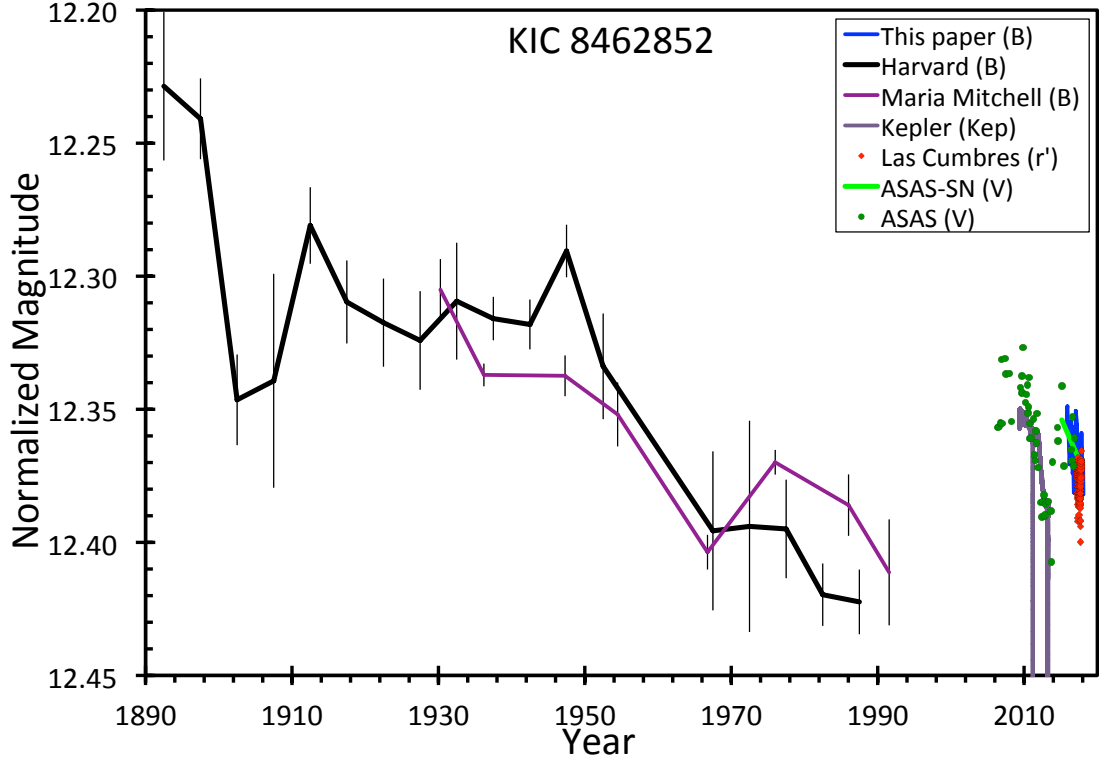}
    \caption{Combined light curve from 1890 to 2018.  The light curves from seven sources have been overplotted with vertical offsets so as to display the century-long ups and downs of KIC 8462852.  In particular, care has been made to determine and display the relative vertical positioning of the 1890--1991 and 2006--2018 segments.  This is critical in showing that the `baseline' in recent years is 0.124 mag fainter than the brightness in the 1890s, thus showing that the modern infrared observations were taken at a time when dust covered at least 12.4\% of the starlight.  The issue is that this near-periastron dust must have emitted a significantly detectable amount of infrared light, with such not being seen by many observers.  Our B-band light curve is the blue line, seen tightly compressed in the middle of the far right hand side.  The Harvard B-band light curve is the black line from upper left to the lower right, and this is correctly placed vertically to be calibrated with the light curve from this paper.  The Maria Mitchell light curve from archival plates is the decadal weighted average, vertically shifted to most closely match the Harvard archival data.  The Kepler light curve is shown as a grey curve in the middle of the right hand side, with the deep dips extending below the bottom of the plot.  The Las Cumbres r'-band light curve is red dots, shifted vertically to match our light curve, and it well covers the blue line from the light curve from this paper.  The ASAS-SN V-band light curve is represented by a light-green line that closely overlaps our blue curve in the far right of the graph.  The ASAS light curve (dark-green dots) has been combined to ten days bins, and then vertically shifted to optimally match our light curve, with this tying the cross calibration back to the {\it Kepler} light curve.}
\end{figure*}

\begin{figure*}
	\includegraphics[width=\textwidth]{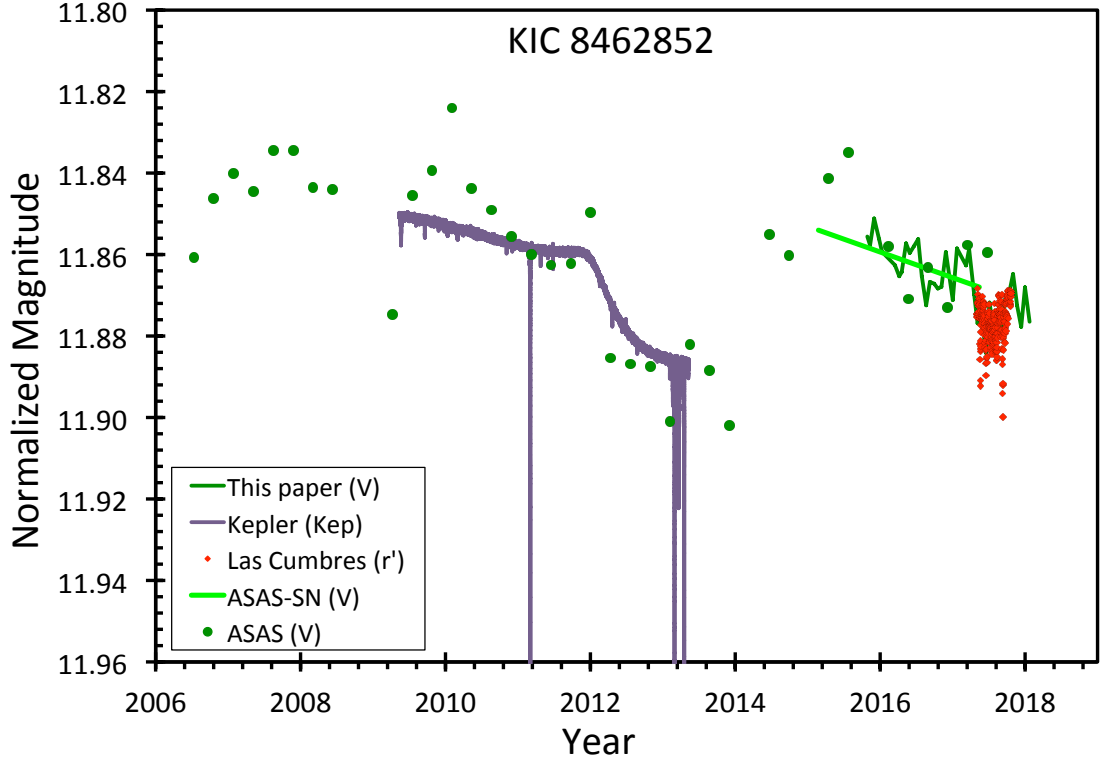}
    \caption{Blow-up of the 2006--2018 lightcurve.  The symbols are described in the legend as well as in the caption for Figure 2.  (Our light curve is represented with the V-band data as a dark green line, with the different band from Figure 2 because the V-band is most comparable with the other light curves in this figure.)  The vertical offsets between the light curves were established in Figure 2.  We see a dip lasting 2.0 years centered on 2013, with the big array of deep {\it Kepler} dips at the bottom of the dip.  And our light curve from 2015.75 to 2018.18 shows a steady decline, with three dips with duration 120--180 days, the deepest of which has superposed the short-duration dips of the {\it Elsie} family.  Within dust models, the secular dimming is where a long string of dust clouds, stretched along some Keplerian orbit, is slowly getting thick for the parts passing over the star, while the dips come from somewhat thicker clumps of varying size.}
\end{figure*}

\begin{figure*}
	\includegraphics[width=\textwidth]{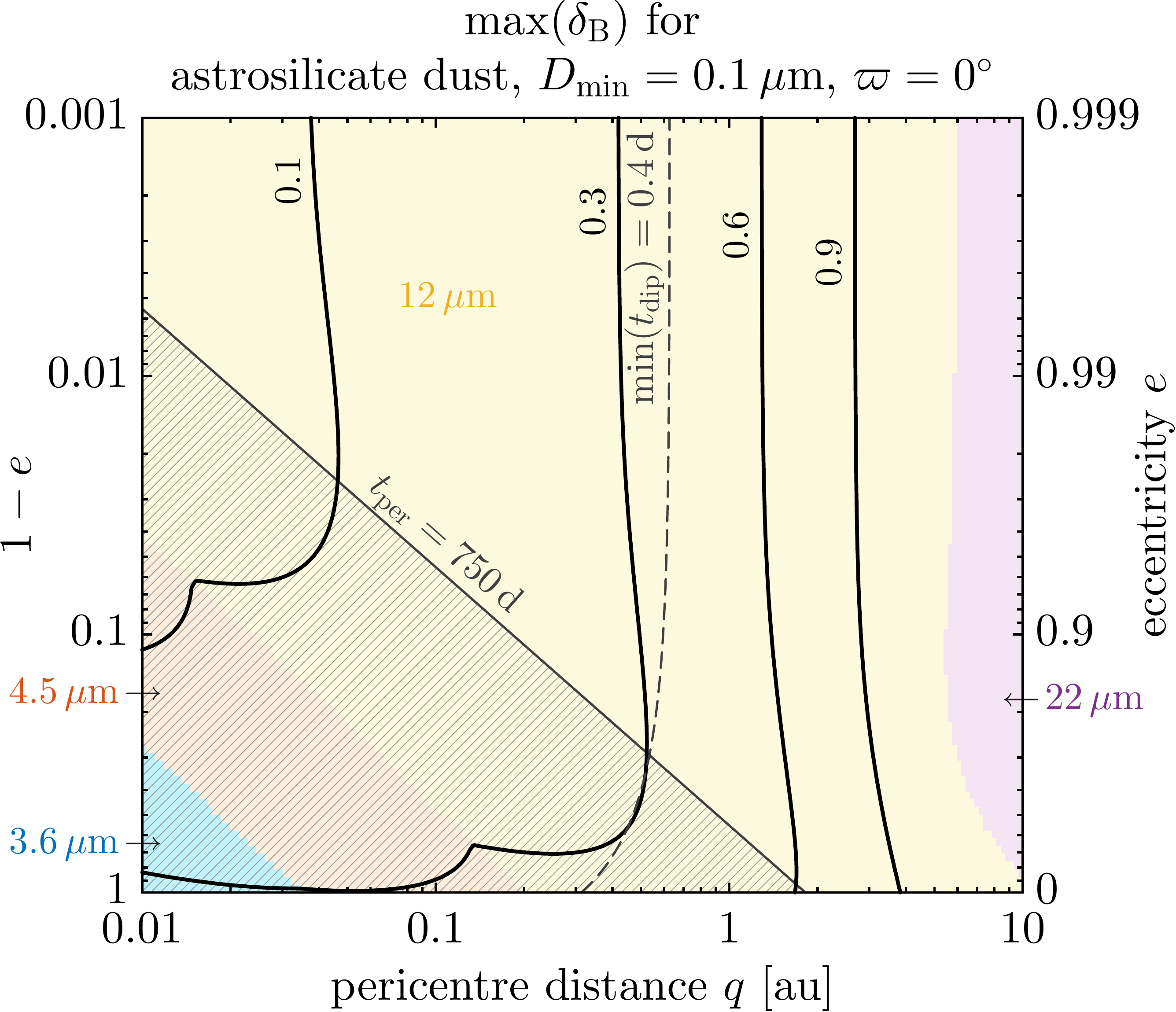}
    \caption{The constraints from the general dust model.  This model has all the occulting dust in a single Keplerian orbit with high eccentricity, with strings of clouds all along the orbit and passing at some point in front of the star.  This orbit is quantified by the distance between the periastron (pericentre) and the star, $q$, the orbital eccentricity $e$, and the longitude of the occulting dust, $\varpi$, equaling zero to show that the occulting dust is at the periastron.  The dust is composed of astrosilicates, with a power law size distribution extending down to 0.1 micron in size.  The nearly-vertical dashed curve is the limit for the cases that allows a dip duration as short as 0.4 days, like for one of the {\it Kepler} dips.  To satisfy this limit, the allowed region is to the left of this dashed curve.  The nearly-vertical solid black curves are for various cases where the dip depth, $max(\delta_B)$, have values of 0.1 (i.e., a 10\% dip), 0.3, 0.6, and 0.9 mag.  The point of this plot is that our limit ($A_B= 0.124 \pm 0.013$ mag) is near the 0.1 curve and always far to the left of the duration limit dashed curve.  With this, the model can produce a 12.4\% dip in the blue for a inner edge of the orbit at around 0.05 AU, all while not violating the limits on either the infrared excess or the dip duration.  That is, our limit is easily compatible with the dust model.  The positioning of the near-vertical curves varies somewhat with the value of $\varpi$, but the same relations still are the same, for example with the dashed curve lying closely on top of the 0.3 dip-depth curve.  The diagonal line labeled `750 d' is the limit imposed by the lack of periodicity in the {\it Kepler} deep dips, with the slant-striped region under the diagonal line being forbidden.  The colored regions display the infrared wavelength for which the most restrictive measure is made.  For all the allowed cases, the most restrictive observation is at 12 microns.}
\end{figure*}

\begin{table*}
	\centering
	\caption{Journal of observations}
	\begin{tabular}{llllllllll} 
		\hline
		Observers & Observatory & Telescope & $N_{mags}$ & $N_{nights}$ & Filters & B Offset & V Offset & R Offset & I Offset\\
		\hline
Coker	&	Sonoita Research Obs. 	&	50-cm, Newt., f/4	&	124	&	69	&	BV	&	-0.035	&	-0.023	&	$\ldots$	&	$\ldots$	\\
Dubois, Logie, 	&	AstroLAB 	&	68.4-cm, Newt., f/4.1	&	3832	&	427	&	BVR	&	-0.016	&	0.001	&	$\equiv$0	&	$\ldots$	\\
~~Rau, \& 	&  ~~~IRIS Obs.	&	&	&	&	&	&	&	&	\\
~~Vanaverbeke 	&	&	&	&	&	&	&	&	&	\\
Dvorak	&	Rolling Hills Obs.	&	35-cm, Cat., f/10	&	193	&	193	&	BV	&	-0.011	&	0.010	&	$\ldots$	&	$\ldots$	\\
Erdelyi	&	KSE Obs.	&	30-cm, SC, f/10	&	307	&	11	&	V	&	$\ldots$	&	0.021	&	$\ldots$	&	$\ldots$	\\
Graham	&	Spica Obs.	&	20-cm, SC, f/10	&	3089	&	58	&	BVRI	&	-0.021	&	0.000	&	-0.040	&	-0.002	\\
Hall	&	Angel Peaks Obs.	&	35-cm, SC, f/10	&	2612	&	63	&	BV	&	-0.040	&	-0.012	&	$\ldots$	&	$\ldots$	\\
Harris	&	Bar J Obs.	&	40-cm, SC, f/8	&	1594	&	15	&	V	&	$\ldots$	&	-0.012	&	$\ldots$	&	$\ldots$	\\
James	&	James Obs.	&	25-cm, SC, f/6.3	&	1640	&	763	&	BVRI	&	-0.010	&	0.019	&	-0.033	&	0.008	\\
Johnston	&	Karen Obs.	&	28-cm, SC, f/7.5	&	2545	&	21	&	BV	&	$\ldots$	&	0.005	&	$\ldots$	&	$\ldots$	\\
Oksanen	&	Hankasalmi Obs.	&	40-cm, RC, f/8.4	&	1135	&	174	&	BVI	&	$\equiv$0	&	$\equiv$0	&	$\ldots$	&	$\equiv$0	\\
Ott	&	Ott Obs.	&	51-cm, Newt., f/4.0	&	1689	&	30	&	VI	&	$\ldots$	&	0.015	&	$\ldots$	&	-0.031	\\
Schaefer, 	&	Highland Road  &	51-cm, RC, f/8.1	&	416	&	42	&	BVR	&	-0.192	&	0.054	&	-0.344	&	$\ldots$	\\
~~Ellis, 	&  ~~~Park Obs.	&	&	&	&	&	&	&	&	\\
~~Nugent, \&	& ~~~(HRPO)	&	&	&	&	&	&	&	&	\\
~~Bentley 	&	&	&	&	&	&	&	&	&	\\
		\hline
	\end{tabular}
\end{table*}

\begin{table}
	\centering
	\caption{1846 Nightly averaged magnitudes (full table is on-line only)}
	\begin{tabular}{lllll} 
		\hline
		Band & $\langle$JD$\rangle$ & Observer & $N_{mags}$ & $\langle$mag$\rangle$\\
		\hline
B	&	2457314	&	Oksanen	&	4	&	12.363	$\pm$	0.003	\\
B	&	2457318	&	Graham	&	6	&	12.394	$\pm$	0.012	\\
B	&	2457320	&	Oksanen	&	5	&	12.360	$\pm$	0.002	\\
B	&	2457322	&	Graham	&	20	&	12.374	$\pm$	0.003	\\
B	&	2457322	&	AstroLAB	&	56	&	12.370	$\pm$	0.001	\\
$\ldots$  &   &   &   &   \\
I	&	2458097	&	James	&	1	&	11.169	$\pm$	0.020	\\
I	&	2458127	&	Oksanen	&	5	&	11.162	$\pm$	0.001	\\
I	&	2458136	&	Oksanen	&	5	&	11.165	$\pm$	0.003	\\
I	&	2458141	&	Oksanen	&	5	&	11.159	$\pm$	0.002	\\
I	&	2458155	&	Oksanen	&	5	&	11.160	$\pm$	0.001	\\
		\hline
	\end{tabular}
\end{table}

\begin{table}
	\centering
	\caption{BVRI Light Curve from 2015.75 to 2018.18 with 20-day resolution}
	\begin{tabular}{llll} 
		\hline
		Band    &     $\langle$JD$\rangle$     & $N_{nights}$   &   $\langle$$\langle$mag$\rangle$$\rangle$\\
		\hline
B	&	2457323.9	&	12	&	12.357	$\pm$	0.002	\\
B	&	2457337.8	&	18	&	12.351	$\pm$	0.005	\\
B	&	2457358.9	&	9	&	12.350	$\pm$	0.002	\\
B	&	2457383.3	&	3	&	12.357	$\pm$	0.003	\\
B	&	2457398.3	&	4	&	12.356	$\pm$	0.003	\\
B	&	2457464.2	&	6	&	12.372	$\pm$	0.007	\\
B	&	2457484.5	&	7	&	12.368	$\pm$	0.005	\\
B	&	2457498.0	&	20	&	12.368	$\pm$	0.003	\\
B	&	2457520.0	&	15	&	12.362	$\pm$	0.003	\\
B	&	2457540.1	&	11	&	12.363	$\pm$	0.007	\\
B	&	2457559.0	&	20	&	12.362	$\pm$	0.006	\\
B	&	2457580.6	&	20	&	12.360	$\pm$	0.004	\\
B	&	2457599.7	&	21	&	12.377	$\pm$	0.009	\\
B	&	2457619.5	&	14	&	12.375	$\pm$	0.005	\\
B	&	2457640.1	&	19	&	12.374	$\pm$	0.002	\\
B	&	2457661.0	&	34	&	12.374	$\pm$	0.002	\\
B	&	2457679.6	&	23	&	12.380	$\pm$	0.003	\\
B	&	2457698.2	&	14	&	12.380	$\pm$	0.006	\\
B	&	2457720.5	&	20	&	12.366	$\pm$	0.004	\\
B	&	2457738.5	&	5	&	12.374	$\pm$	0.014	\\
B	&	2457755.2	&	3	&	12.364	$\pm$	0.005	\\
B	&	2457776.4	&	6	&	12.352	$\pm$	0.008	\\
B	&	2457798.3	&	3	&	12.361	$\pm$	0.020	\\
B	&	2457823.7	&	7	&	12.373	$\pm$	0.011	\\
B	&	2457837.2	&	12	&	12.357	$\pm$	0.006	\\
B	&	2457863.2	&	10	&	12.368	$\pm$	0.009	\\
B	&	2457880.7	&	15	&	12.374	$\pm$	0.006	\\
B	&	2457897.8	&	19	&	12.377	$\pm$	0.006	\\
B	&	2457921.8	&	18	&	12.380	$\pm$	0.005	\\
B	&	2457938.8	&	13	&	12.370	$\pm$	0.005	\\
B	&	2457962.6	&	5	&	12.384	$\pm$	0.005	\\
B	&	2457980.6	&	10	&	12.388	$\pm$	0.004	\\
B	&	2458000.0	&	23	&	12.381	$\pm$	0.004	\\
B	&	2458019.5	&	24	&	12.376	$\pm$	0.002	\\
B	&	2458040.6	&	18	&	12.367	$\pm$	0.005	\\
B	&	2458059.6	&	29	&	12.365	$\pm$	0.004	\\
B	&	2458079.8	&	24	&	12.373	$\pm$	0.004	\\
B	&	2458100.5	&	9	&	12.379	$\pm$	0.008	\\
B	&	2458119.0	&	10	&	12.381	$\pm$	0.003	\\
B	&	2458141.9	&	8	&	12.370	$\pm$	0.003	\\
B	&	2458162.6	&	6	&	12.367	$\pm$	0.016	\\
B	&	2458176.7	&	4	&	12.379	$\pm$	0.013	\\
V	&	2457295.4	&	1	&	11.816	$\pm$	0.010	\\
V	&	2457324.3	&	17	&	11.849	$\pm$	0.004	\\
V	&	2457337.9	&	19	&	11.849	$\pm$	0.003	\\
V	&	2457358.9	&	9	&	11.843	$\pm$	0.002	\\
V	&	2457383.3	&	3	&	11.847	$\pm$	0.002	\\
V	&	2457398.3	&	4	&	11.855	$\pm$	0.002	\\
V	&	2457463.5	&	5	&	11.856	$\pm$	0.005	\\
V	&	2457485.4	&	9	&	11.861	$\pm$	0.005	\\
V	&	2457497.8	&	21	&	11.860	$\pm$	0.003	\\
V	&	2457519.6	&	21	&	11.851	$\pm$	0.002	\\
V	&	2457538.4	&	13	&	11.853	$\pm$	0.005	\\
V	&	2457558.9	&	17	&	11.852	$\pm$	0.003	\\
V	&	2457580.9	&	19	&	11.850	$\pm$	0.003	\\
V	&	2457599.0	&	22	&	11.857	$\pm$	0.003	\\
V	&	2457620.8	&	18	&	11.864	$\pm$	0.003	\\
V	&	2457640.0	&	25	&	11.862	$\pm$	0.002	\\
V	&	2457661.4	&	36	&	11.862	$\pm$	0.002	\\
V	&	2457679.4	&	25	&	11.863	$\pm$	0.003	\\
V	&	2457698.0	&	15	&	11.861	$\pm$	0.004	\\
V	&	2457720.4	&	21	&	11.853	$\pm$	0.004	\\
V	&	2457736.9	&	5	&	11.863	$\pm$	0.015	\\
V	&	2457755.2	&	3	&	11.862	$\pm$	0.004	\\
		\hline
	\end{tabular}
\end{table}

\begin{table}
	\centering
	\contcaption{BVRI Light Curve from 2015.75 to 2018.18}
	\label{tab:continued}
	\begin{tabular}{llll} 
		\hline
		Band    &     $\langle$JD$\rangle$     & $N_{nights}$   &   $\langle$$\langle$mag$\rangle$$\rangle$\\
		\hline
V	&	2457776.4	&	6	&	11.849	$\pm$	0.004	\\
V	&	2457798.3	&	3	&	11.870	$\pm$	0.009	\\
V	&	2457822.1	&	12	&	11.854	$\pm$	0.004	\\
V	&	2457837.4	&	13	&	11.851	$\pm$	0.004	\\
V	&	2457863.0	&	17	&	11.862	$\pm$	0.006	\\
V	&	2457879.9	&	18	&	11.865	$\pm$	0.004	\\
V	&	2457898.3	&	27	&	11.867	$\pm$	0.003	\\
V	&	2457922.0	&	24	&	11.877	$\pm$	0.003	\\
V	&	2457940.3	&	21	&	11.864	$\pm$	0.003	\\
V	&	2457959.4	&	7	&	11.873	$\pm$	0.004	\\
V	&	2457980.7	&	11	&	11.872	$\pm$	0.003	\\
V	&	2457999.9	&	32	&	11.873	$\pm$	0.002	\\
V	&	2458019.2	&	33	&	11.865	$\pm$	0.002	\\
V	&	2458041.7	&	25	&	11.861	$\pm$	0.003	\\
V	&	2458059.1	&	30	&	11.859	$\pm$	0.003	\\
V	&	2458079.4	&	25	&	11.867	$\pm$	0.003	\\
V	&	2458100.1	&	10	&	11.872	$\pm$	0.004	\\
V	&	2458119.0	&	10	&	11.862	$\pm$	0.006	\\
V	&	2458142.6	&	9	&	11.869	$\pm$	0.003	\\
V	&	2458162.9	&	8	&	11.867	$\pm$	0.008	\\
V	&	2458176.4	&	6	&	11.869	$\pm$	0.002	\\
R	&	2457324.5	&	7	&	11.453	$\pm$	0.001	\\
R	&	2457338.6	&	9	&	11.450	$\pm$	0.001	\\
R	&	2457359.3	&	6	&	11.447	$\pm$	0.002	\\
R	&	2457383.3	&	2	&	11.453	$\pm$	0.000	\\
R	&	2457395.8	&	2	&	11.456	$\pm$	0.004	\\
R	&	2457463.3	&	5	&	11.447	$\pm$	0.008	\\
R	&	2457481.0	&	4	&	11.466	$\pm$	0.011	\\
R	&	2457497.3	&	10	&	11.431	$\pm$	0.010	\\
R	&	2457519.5	&	8	&	11.441	$\pm$	0.005	\\
R	&	2457540.3	&	6	&	11.456	$\pm$	0.004	\\
R	&	2457557.6	&	11	&	11.445	$\pm$	0.006	\\
R	&	2457582.2	&	10	&	11.445	$\pm$	0.005	\\
R	&	2457596.7	&	6	&	11.455	$\pm$	0.002	\\
R	&	2457622.0	&	3	&	11.461	$\pm$	0.003	\\
R	&	2457640.7	&	7	&	11.464	$\pm$	0.002	\\
R	&	2457661.3	&	21	&	11.462	$\pm$	0.001	\\
R	&	2457681.3	&	15	&	11.458	$\pm$	0.004	\\
R	&	2457699.0	&	12	&	11.445	$\pm$	0.007	\\
R	&	2457719.1	&	16	&	11.444	$\pm$	0.004	\\
R	&	2457739.6	&	3	&	11.439	$\pm$	0.010	\\
R	&	2457755.2	&	3	&	11.448	$\pm$	0.004	\\
R	&	2457773.7	&	5	&	11.441	$\pm$	0.004	\\
R	&	2457798.3	&	3	&	11.446	$\pm$	0.007	\\
R	&	2457827.2	&	3	&	11.450	$\pm$	0.005	\\
R	&	2457838.6	&	8	&	11.449	$\pm$	0.003	\\
R	&	2457863.2	&	10	&	11.453	$\pm$	0.004	\\
R	&	2457880.7	&	12	&	11.455	$\pm$	0.008	\\
R	&	2457898.4	&	14	&	11.457	$\pm$	0.003	\\
R	&	2457920.9	&	16	&	11.456	$\pm$	0.003	\\
R	&	2457939.9	&	14	&	11.458	$\pm$	0.003	\\
R	&	2457969.8	&	1	&	11.490	$\pm$	0.010	\\
R	&	2457979.9	&	7	&	11.475	$\pm$	0.011	\\
R	&	2457999.2	&	14	&	11.469	$\pm$	0.004	\\
R	&	2458019.2	&	14	&	11.493	$\pm$	0.015	\\
R	&	2458040.4	&	14	&	11.486	$\pm$	0.014	\\
R	&	2458059.8	&	13	&	11.513	$\pm$	0.021	\\
R	&	2458078.9	&	16	&	11.491	$\pm$	0.019	\\
R	&	2458101.4	&	6	&	11.456	$\pm$	0.006	\\
R	&	2458121.8	&	4	&	11.456	$\pm$	0.007	\\
R	&	2458142.1	&	5	&	11.461	$\pm$	0.004	\\
R	&	2458161.1	&	3	&	11.472	$\pm$	0.002	\\
R	&	2458176.7	&	4	&	11.472	$\pm$	0.003	\\
		\hline
	\end{tabular}
\end{table}

\begin{table}
	\centering
	\contcaption{BVRI Light Curve from 2015.75 to 2018.18}
	\label{tab:continued}
	\begin{tabular}{llll} 
		\hline
		Band    &     $\langle$JD$\rangle$     & $N_{nights}$   &   $\langle$$\langle$mag$\rangle$$\rangle$\\
		\hline
I	&	2457322.9	&	8	&	11.155	$\pm$	0.001	\\
I	&	2457335.8	&	8	&	11.151	$\pm$	0.002	\\
I	&	2457400.8	&	2	&	11.153	$\pm$	0.002	\\
I	&	2457464.3	&	6	&	11.158	$\pm$	0.003	\\
I	&	2457483.2	&	5	&	11.154	$\pm$	0.001	\\
I	&	2457496.7	&	15	&	11.157	$\pm$	0.001	\\
I	&	2457518.1	&	10	&	11.152	$\pm$	0.001	\\
I	&	2457538.5	&	5	&	11.148	$\pm$	0.002	\\
I	&	2457556.4	&	7	&	11.152	$\pm$	0.001	\\
I	&	2457582.3	&	6	&	11.156	$\pm$	0.003	\\
I	&	2457597.4	&	3	&	11.152	$\pm$	0.004	\\
I	&	2457616.8	&	3	&	11.157	$\pm$	0.007	\\
I	&	2457640.5	&	12	&	11.158	$\pm$	0.001	\\
I	&	2457661.1	&	19	&	11.160	$\pm$	0.001	\\
I	&	2457678.5	&	11	&	11.157	$\pm$	0.003	\\
I	&	2457698.8	&	7	&	11.157	$\pm$	0.003	\\
I	&	2457717.9	&	11	&	11.153	$\pm$	0.003	\\
I	&	2457733.2	&	1	&	11.138	$\pm$	0.010	\\
I	&	2457827.3	&	4	&	11.154	$\pm$	0.002	\\
I	&	2457840.2	&	8	&	11.154	$\pm$	0.002	\\
I	&	2457863.2	&	10	&	11.157	$\pm$	0.004	\\
I	&	2457880.4	&	12	&	11.158	$\pm$	0.004	\\
I	&	2457898.0	&	11	&	11.159	$\pm$	0.003	\\
I	&	2457920.3	&	11	&	11.162	$\pm$	0.002	\\
I	&	2457937.6	&	6	&	11.160	$\pm$	0.002	\\
I	&	2457969.8	&	1	&	11.168	$\pm$	0.010	\\
I	&	2457979.1	&	3	&	11.163	$\pm$	0.003	\\
I	&	2458000.2	&	16	&	11.162	$\pm$	0.002	\\
I	&	2458020.7	&	14	&	11.157	$\pm$	0.002	\\
I	&	2458042.1	&	11	&	11.157	$\pm$	0.003	\\
I	&	2458057.6	&	8	&	11.165	$\pm$	0.005	\\
I	&	2458079.3	&	9	&	11.154	$\pm$	0.004	\\
I	&	2458096.6	&	1	&	11.177	$\pm$	0.010	\\
I	&	2458127.2	&	1	&	11.162	$\pm$	0.010	\\
I	&	2458138.7	&	2	&	11.162	$\pm$	0.002	\\
I	&	2458155.2	&	1	&	11.160	$\pm$	0.010	\\
		\hline
	\end{tabular}
\end{table}

\begin{table*}
	\centering
	\caption{Slopes and amplitudes as a function of color over five sets of time intervals}
	\begin{tabular}{llllll} 
		\hline
Slope or Amplitude & $JD-2450000$ range & B-band & V-band & R-band & I-band \\
		\hline
Slope overall (\% per year)	&	7295 to 8183	&	0.97	$\pm$	0.09	&	0.94	$\pm$	0.07	&	0.69	$\pm$	0.09	&	0.39	$\pm$	0.09	\\
Slope around Elsie (\% per year)	&	7770 to 7990	&	5.68	$\pm$	0.93	&	3.77	$\pm$	0.68	&	4.10	$\pm$	0.70	&	2.44	$\pm$	0.62	\\
Amplitude, Elsie Group to Post-Angkor (mag)	&	7890 to 8010 vs. 8030 to 8090	&	0.016	$\pm$	0.004	&	0.009	$\pm$	0.002	&	0.003	$\pm$	0.003	&	0.002	$\pm$	0.003	\\
Amplitude, 1st peak to 2nd dip (mag)	&	7510 to 7590 vs. 7610 to 7710	&	0.016	$\pm$	0.003	&	0.011	$\pm$	0.002	&	0.014	$\pm$	0.003	&	0.007	$\pm$	0.002	\\
Amplitude, start to end (mag)	&	7295 to 7370 vs. 8110 to 8183	&	0.023	$\pm$	0.003	&	0.022	$\pm$	0.003	&	0.018	$\pm$	0.003	&	0.009	$\pm$	0.003	\\
		\hline
	\end{tabular}
\end{table*}

\bsp	
\label{lastpage}
\end{document}